\documentclass[aps,prl,showpacs,twocolumn]{revtex4}
\usepackage{graphicx}
\usepackage{amsmath}

\begin{document}

\title{Thermal conductance of Andreev interferometers}
\author{Z. Jiang and V. Chandrasekhar}
\affiliation{Department of Physics and Astronomy, Northwestern University, Evanston, IL 60208, USA}
\date{\today}
\pacs{74.25.Fy, 74.45.+c, 73.23.-b}

\begin{abstract}
We calculate the thermal conductance $G^T$ of diffusive Andreev interferometers, which are hybrid loops with one superconducting arm and one normal-metal arm.  The presence of the superconductor suppresses $G^T$; however, unlike a conventional superconductor, $G^T/G^T_N$ does not vanish as the temperature $T\rightarrow0$, but saturates at a finite value that depends on the resistance of the normal-superconducting interfaces, and their distance from the path of the temperature gradient.  The reduction of $G^T$ is determined primarily by the suppression of the density of states in the proximity-coupled normal metal along the path of the temperature gradient.  $G^T$ is also a strongly nonlinear function of the thermal current, as found in recent experiments. 
\end{abstract}

\maketitle

The thermal transport properties of proximity-coupled normal metals have attracted interest recently \cite{heikkila,bezuglyi,zhao}, spurred by recent measurements of the thermopower $S_A$ of Andreev interferometers \cite{eom, dikin1, parsons}, which are loops in which one arm is fabricated from a superconductor (S), and the second arm from a normal metal (N).  In these experiments, which include some of our own, the thermopower was found to oscillate periodically as a function of applied magnetic flux, with a fundamental period corresponding to $\Phi_0=h/2e$.  In spite of some recent theoretical attempts \cite{heikkila}, these experimental results have not been explained satisfactorily. 

Given this problem, we have more recently turned our attention to measurements of the thermal conductance $G^T$ of proximity-coupled systems \cite{jiang}, which are more easily calculated in the quasiclassical approximation.  Bezuglyi and Vinokur \cite{bezuglyi} recently discussed $G^T$ for a NSN sandwich where the size of S is smaller than the coherence length.  They found a strong suppression of $G^T$ with decreasing temperature $T$ and oscillations with the applied flux $\Phi$.  However, this behavior was determined primarily by the modification of the thermal conductance of the superconducting component of the sandwich arising from proximity to the normal metal, an \textit{inverse} proximity effect.    In this Letter, we calculate $G^T$ arising from a normal-metal proximity effect for two types of Andreev interferometers relevant to the experiments, the `house' geometry, and the `parallelogram' geometry shown in Fig. 1.  Our intent is to determine the effect of sample geometry and finite NS interface resistance on $G^T$.  As before, we find that the normalized thermal conductance $G^T/G^T_N$ of these devices decreases as the temperature is reduced below the transition temperature of the superconductor \cite{bezuglyi}, but saturates at a finite value that depends on the transparency of the NS interface and the distance of the NS interfaces from the path of the thermal current.  The thermal conductance is highly nonlinear, varying with the heat current $I^T$ approximately as $\sqrt{I^T}$ for intermediate values of $I^T$.  Finally, $G^T$ oscillates periodically with an applied magnetic flux with a fundamental period of $h/2e$ and an amplitude corresponding to a few percent of $G^T_N$, with the amplitude of the oscillations in the `house' geometry being in general larger than those in the `parallelogram' geometry.  For both the `house' and the `parallelogram' geometries, the oscillations in $G^T$ are symmetric in the applied flux.  The suppression in $G^T$ is directly related to the suppression in the density of states in the path of the thermal current.
\begin{figure}[b]
\includegraphics[width=8.0cm]{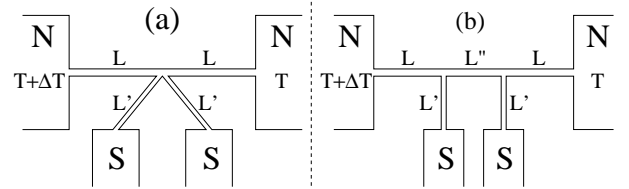}
\caption{Schematic of Andreev interferometers with two different geometries: (a) `house' and (b) `parallelogram'.}
\end{figure}

Our starting point is the kinetic equations for the non-equilibrium distribution functions in the diffusive limit \cite{heikkila,larkin,chandrasekhar,belzig}:
\begin{subequations} \label{eqn1}
\vspace{-0.2cm}
\begin{align}
\partial_{\vec{R}} \left[M_{00}(\partial_{\vec{R}}h_L) + Q h_T + M_{30}(\partial_{\vec{R}}h_T) \right]&=0, \label{eqn1a} \\
\partial_{\vec{R}} \left[M_{33}(\partial_{\vec{R}}h_T) + Q h_L + M_{03}(\partial_{\vec{R}}h_L) \right]&=0.
\label{eqn1b}
\end{align}
\end{subequations}     
Here $h_L$ and $h_T$ are the components of the full matrix distribution function $\tilde{h}=h_L \tau^0+h_T \tau^3$, where $\tau^0$ is the identity matrix, and $\tau^3$ has the form of the Pauli spin matrix $\sigma_z$.  In the normal-metal and superconducting reservoirs, $h_L$ and $h_T$ have the equilibrium forms
\begin{equation}
h_{L,T}=\frac{1}{2}\left[\tanh\left(\frac{E+eV}{2 k_B T}\right) \pm \tanh\left(\frac{E-eV}{2 k_B T}\right)\right],
\label{eqn2}
\end{equation}
where $V$ is the voltage applied to the reservoir.
$M_{00}$, $M_{33}$, $M_{03}$ and $M_{30}$ are given by
%\begin{subequations} \label{eqn7.14}
\begin{align}
M_{00}&= \frac{1}{2}\left[1 + \cosh \theta \cosh \theta^* - \sinh \theta \sinh \theta^* \cosh(2 \Im (\chi))\right], \nonumber   \\
M_{33}&= \frac{1}{2}\left[1 + \cosh \theta \cosh \theta^* + \sinh \theta \sinh \theta^* \cosh( 2\Im(\chi))\right],  \nonumber  \\
M_{03}&= \frac{1}{2}\sinh\theta \sinh \theta^* \sinh(2 \Im(\chi))), \nonumber  \\
%\intertext{and}
M_{30}&= -\frac{1}{2}\sinh\theta \sinh \theta^* \sinh(2\Im(\chi)). \label{eqn3} 
\end{align}
%\end{subequations}    
where $\theta$ and $\chi$ are the solutions of the Usadel equations
\begin{subequations} \label{eqn4}
\begin{align}
&D\partial_{\vec{R}}j_s(E,\vec{R}) = 0,  \label{eqn3a} \\
&D  \partial_{\vec{R}}^2 \theta -\frac{D}{2} \sinh2\theta \;(\partial_{\vec{R}} \chi)^2 + 2Ei\sinh\theta = 0. \label{eqn3b}
\end{align}
\end{subequations}
In writing these equations, we have assumed that there is no superconducting gap in the proximity-coupled normal metal, and that the phase $\chi$ is real in the S reservoirs.  $D$ is the electronic diffusion coefficient, and $j_s$ is the spectral supercurrent given by
\begin{equation}
j_s(E,\vec{R})=\sinh^2\theta(E,\vec{R}) \partial_{\vec{R}}\chi(E,\vec{R}),
\label{eqn5}
\end{equation}
and $Q$ in Eqn. (1) is related to $j_s$ by $Q(E, \vec{R}) = - \Im(j_s(E,\vec{R}))$.    

To close the set of equations, we must specify the boundary conditions at the NS interfaces and at the nodes where multiple one-dimensional normal-metal segments meet.  In the limit of low transparent NS interfaces, the boundary conditions can be written in the form \cite{belzig,kuprianov}
\begin{subequations}\label{eqn6}
\vspace{-0.25cm}
\begin{align}
r &\sinh \theta_1 (\partial_{\vec{R}} \chi_1) =\sinh\theta_2 \sin(\chi_2 - \chi_1), \label{eqn6a} \\
r & \left[\partial_{\vec{R}} \theta_1 +  i \sinh \theta_1 \cosh \theta_1(\partial_{\vec{R}} \chi_1)\right]=  \nonumber \\
& \cosh \theta_1 \sinh \theta_2 e^{i(\chi_2 - \chi_1)}-\sinh \theta_1 \cosh \theta_2 . \label{eqn6b}
\end{align}
\end{subequations}
where the subscript 1 (2) refers to the normal metal (superconductor).  $r=R_b/R_N$ is a parameter that characterizes the transparency of the interface, where $R_b$ is the resistance of the barrier, and $R_N$ the resistance of the normal metal per unit length.   In the limit of $r=0$, these boundary conditions reduce to the continuity of the $\theta$ function and the phase $\chi$ across the interface.  At a superconducting reservoir, $\theta$ is given by $\cosh \theta_{S0} = E/\sqrt{E^2-|\Delta|^2}$, where $\Delta$ is the gap in the superconductor.  At a normal reservoir, $\theta_{N0}=0$.  The major contribution to the phase $\chi$ in our simulations is from the applied magnetic flux.  Since the critical current in the superconductor is much larger than that in the proximity-coupled normal metal, we assume that all the phase change occurs across the normal wires, this change being applied symmetrically between the two NS interfaces.  For example, for an applied flux of $\Phi_0/2$ through the area of the Andreev interferometer loop, we set the phases $\chi$ at the two NS interfaces in the Andreev interferometers to be $-\pi /2$ and $\pi/2$.  For a normal reservoir, the absolute value of $\chi$ is meaningless, but the gradient $\partial_x \chi$ must be 0. 

For a node, assuming one-dimensional wires of equal cross-section, the boundary conditions reduce to the condition that the parameter $\theta$ must be continuous at the node, and $\sum \partial_x \theta = 0$, where the sum is over all wires emanating from the node.  Similar boundary conditions apply to the phase $\chi$.   

In terms of these parameters, the thermal current is given by 
\begin{equation}
I^T=N_0D \int dE \; E [M_{00}(\partial_{\vec{R}}h_L) + Q h_T + M_{30}(\partial_{\vec{R}}h_T)],
\label{eqn7}
\end{equation}
where $N_0$ is the density of states at the Fermi energy. It should be noted that the integrand in the equation above is simply $E$ times the term in the brackets in Eqn. (1a), so that Eqn. (1a) is simply a statement of the conservation of spectral thermal current.

To determine the thermal current for the Andreev interferometers, one must solve the coupled systems of differential equations (1) and (4) numerically.  In general, there is a finite supercurrent flowing between the NS interfaces in an Andreev interferometer in the presence of a magnetic field, so that the terms involving the supercurrent in the equations above cannot be ignored.  For the geometries shown in Fig. 1, however, some simplifications can be made.  We shall consider first the `house' geometry (Fig. 1(a)), applying a temperature differential $\Delta T$ as shown in the figure, so that one normal reservoir is at a temperature $T$ while the other reservoir is at a temperature $T+\Delta T$.  Since no supercurrent flows in the segments connected to the N reservoirs, the phase $\chi$ in these segments is constant.  Furthermore, applying the boundary conditions for the phase $\chi$ as discussed above, $\chi=0$ at the central node by symmetry (this is also confirmed by the numerical simulations), so that we can take $\chi=0$ in the segments connected to the normal reservoirs.  Under these conditions, the last two terms in Eqn. (1a) vanish.  The first integral of the resulting equation is a constant, which we denote $K_1$.  Integrating a second time from the left N reservoir ($x=0$) to the right N reservoir ($x=2L$), we obtain 
\begin{equation}
h_L(x=2L) - h_L(x=0)= K_1 \int_0^{2L} \frac{1}{M_{00}} dx
\end{equation}   
The left hand side of the equation above is determined by the boundary conditions on the distribution function, which are given by Eqn. (2) with $V=0$ and the temperatures as defined by Fig. 1.  The definite integral on the right hand side can be evaluated numerically by solving the Usadel equations (4).  For a small temperature difference $\Delta T$, in the linear response regime, the left hand side can be expanded in a Taylor's series to give the thermal conductance $G^T=I^T/\Delta T$
\begin{equation}
G^T =-\frac{N_0 D}{2 k_B T^2} \int dE \frac{E^2}{\cosh^2(E/2k_B T)} \left[\int_0^{2L} \frac{1}{M_{00}} dx \right]^{-1}.
\end{equation}

\begin{figure}[b]
\includegraphics[width=8.0cm]{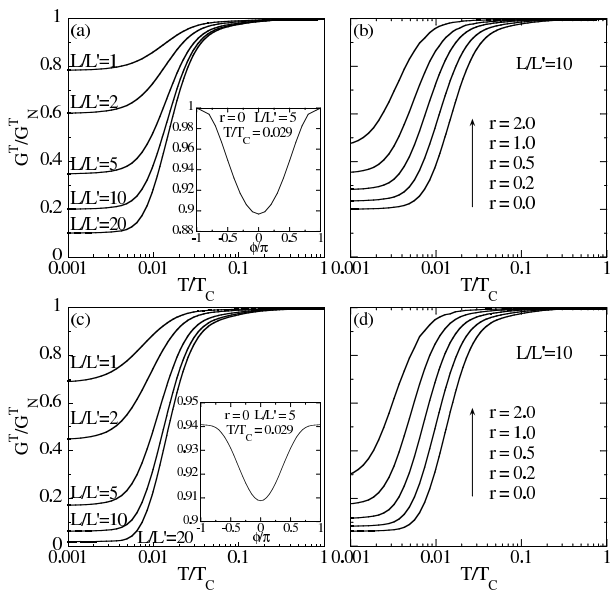}
\caption{$G^T$ of the `house' interferometer ((a) and (b)) and the `parallelogram' interferometer ((c) and (d)) as a function of the normalized temperature $T/T_c$. (a) and (c): for different $L/L'$; (b) and (d): for different $r$. The insets in (a) and (c) show the oscillations of $G_T$ as a function of the phase $\phi$.}
\end{figure}

Figure 2(a) shows $G^T$ of the `house' interferometer as a function of the normalized temperature $T/T_c$, for a sample with a perfectly transparent barrier ($r=0$).  The length $L$ of one normal-metal segment attached to a normal reservoir defines the correlation energy $E_c = \hbar D/L^2$ (see Fig. 1).  The transition temperature $T_c$ of the superconductor is chosen to be 17.18 $E_c/k_B$, corresponding closely to the values in the experiments.  The different curves in Fig. 2(a) correspond to different lengths $L'$ of the normal-metal segments connected to the superconducting reservoirs.  $G^T$ differs appreciably from its normal state value $G^T_N$ only at temperatures below $E_c/k_B$, and saturates as $T\rightarrow0$.  The overall decrease in $G^T/G^T_N$ depends on the length of the segment $L'$, the decrease being the greatest for the smallest value of $L'$.  It should be noted, however, that even for the smallest value of $L'$, $G^T/G^T_N$ does not vanish at the lowest temperatures, in contrast to the case of a pure superconductor \cite{abrikosov}.  

It is well known that the electric conductance $G$ of a proximity-coupled normal metal shows the so-called reentrance behavior \cite{stoof}: as one decreases the temperature below $T_c$, $G$ first increases and then decreases, approaching its normal state value $G_N$ as $T\rightarrow0$. This non-monotonic temperature dependence is a result of the competition between two effects.  One is the penetration of superconducting correlations from the S side into the N side, which increases the value of $G$; the other is the reduction of the density of states of electrons, which reduces the value of $G$.  The reduction in $G^T$, however, is directly associated only with the suppression of the density of states $N(E,x)$ in the proximity-coupled normal metal \cite{dos}.  This leads to a reduction in the number of quasiparticles that can carry a thermal current.  Fig. 3(a) shows the normalized $N(E,x)=\cosh(\Re(\theta)) \cos(\Im(\theta))$ for the `house'  interferometer along the path of the thermal current from $x=0$ to $x=2L$, for $r=0$ and $L/L'=10$.  The suppression is maximum at the node ($x=L$); however, unlike a superconductor, $N(E,x)$ never vanishes below the gap.  Hence, $G^T/G^T_N$ is suppressed, but remains finite as $T\rightarrow0$.  In a normal metal, $G^T_N$ and $G_N$ obey the Wiedemann-Franz law, $G^T_N \propto G_NT$ \cite{wiedemann}. In a proximity-coupled normal-metal, as in a superconductor, the Wiedemann-Franz law is violated.  Physically, this violation results from the enhanced conductance associated with the superconducting correlations induced in the normal metal, which do not carry any thermal current.

\begin{figure}[b]
\includegraphics[width=8.0cm]{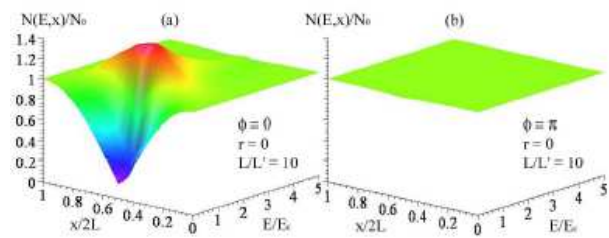}
\caption{Density of states $N(E,x)$ for the `house' interferometer along the path of the thermal current for $r=0$ and $L/L'=10$. (a) $\phi=0$; (b) $\phi=\pi$.}
\end{figure}

Any parameter that affects  $N(E,x)$ will modify $G^T$.  One parameter is clearly $L'$, the distance between the S reservoir and the path of the thermal current, as shown in Fig. 2(a).  A second parameter is the interface resistance $r$.  A larger value of $r$ would reduce the proximity effect in the normal metal, leading to a smaller change in $G^T$.  Figure 2(b) shows $G^T$ as a function of $T/T_c$ for the case $L/L'=10$, for a number of different values of $r$.  As expected, the smaller the value of $r$, the greater the reduction in $G^T$.   As both $L'$ and $r$ affect $G^T$, understanding the dependence of $G^T$ on $L'$ and $r$ is crucial for a quantitative comparison to the experimental results.

The third parameter that affects $N(E,x)$ is the applied magnetic flux.  Figure 3(a) shows $N(E,x)$ for zero phase difference $\phi=0$ between the two S reservoirs; Fig. 3(b) shows $N(E,x)$ for $\phi=\pi$, corresponding to a flux $\Phi_0/2$ through the interferometer loop.  For $\Phi_0/2$, $N(E,x)$ is essentially flat along the entire normal-metal segment at a value corresponding to the normal-metal density of states $N_0$.  Consequently, $G^T$ regains its normal state value $G^T_N$ at $\phi=\pi$, so that $G^T$ shows full scale oscillations as a function of magnetic flux.  This can be seen in the inset of Fig. 2(a), which shows $G^T$ as a function of $\phi$ for $r=0$ and $L/L'=5$ at a temperature of $T/T_c=0.029$.  The observation of full-scale oscillations is directly related to the symmetry of the `house'  interferometer.  We note that the oscillations are symmetric with respect to magnetic field.  If the central node were not exactly midway between the two S reservoirs, there would still be a suppression of $N(E,x)$ at the node even for $\phi=\pi$, and hence $G^T$ would never approach $G^T_N$ for any value of $\phi$.  Hence the amplitude of the oscillations are reduced if the geometry of the `house' interferometer is not symmetric.

We now turn to the `parallelogram' interferometer, shown in Fig. 1(b).  In this case, in addition to the two normal-metal segments of length $L$ connected to the N reservoirs, we also have a normal-metal segment of length $L''$ in the path of the thermal current.  We use a typical experimental value of $L''/L=0.66$ for all the simulations in this Letter. Since this segment lies between the S reservoirs, a supercurrent may flow in this segment, so that the last two terms in the square brackets in Eqn. (1a) cannot in general be ignored.  However, since $V$ in the S reservoirs is 0, $h_T=0$ at the S reservoirs, so that the terms involving $h_T$ are small and can be ignored.  This assumption is also supported by detailed numerical simulations \cite{heikkila}.  Under these conditions, the equation for $G^T$ is again given by an expression similar to Eqn. (9), except that the integral of $1/M_{00}$ is over the three normal-metal segments along which the thermal current flows.

\begin{figure}[t]
\includegraphics[width=6cm]{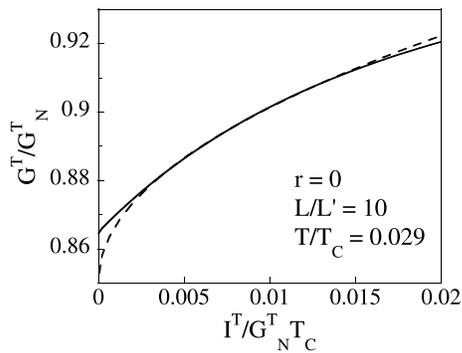}
\caption{Solid lines represent the strongly nonlinear $G^T$ as a function of the thermal current $I^T$ for the `house' interferometer. Dashed lines illustrate the $\sqrt{I^T}$ dependence of $G^T$ at intermediate values of the thermal current.}
\end{figure}

Figures 2(c) and 2(d) show $G^T$ for the `parallelogram' interferometer as a function of $L/L'$ and $r$ respectively.  The dependence on $L/L'$ and $r$ is qualitatively similar to that of the `house' interferometer.  However, unlike the case for the `house' interferometer, the oscillations of $G^T$ as a function of the phase $\phi$ (shown in the inset of Fig. 2(c)) do not reach their full scale amplitude, i.e., $G^T$ is less than $G^T_N$ for all values of $\phi$, because there would always be a suppression in the density of states around the two nodes for this geometry. We also note that the oscillations are symmetric with respect to $\phi$ (and hence, with respect to an applied flux); if the last two terms in Eqn. (1a) were not negligible, we would also have small antisymmetric contributions to $G^T$, since both $Q$ and $M_{30}$ are antisymmetric in the applied flux.    

In experiments on the thermal conductance of Andreev interferometers, $G^T$ was found to be a strongly nonlinear function of the thermal current $I^T$ \cite{dikin2,jiang}.  In order to investigate the dependence of $G^T$ on $I^T$, we do not take the linear response limit of Eqn. (8), but numerically calculate $K_1$ for specific values of $\Delta T$ across the interferometer. $G^T$ is still defined by the ratio $I^T/\Delta T$, but it is no longer given by the linear response result Eqn. (9).  Figure 4 shows the calculated $G^T$ as a function of $I^T$ for the `house' interferometer (the `parallelogram' interferometer shows similar behavior). If $G^T$ were linear for small $I^T$, we would expect a curve which had zero slope at $I^T=0$, which is clearly not seen in the figure.  At intermediate values of the thermal current, $G^T$ varies as $\sqrt{I^T}$, shown by the dashed line in the figure.  This nonlinear dependence of $G^T$ is also observed in the experiments and presents a problem in experimentally defining the linear-response thermal conductance of the sample \cite{jiang}.    

We thank W. Belzig for useful discussions.  This work is supported by the NSF under grant number DMR-0201530.

\end{document}